\documentclass[fleqn,usenatbib]{mnras}

\usepackage{newtxtext,newtxmath}

\usepackage[T1]{fontenc}

\usepackage{graphicx}	
\usepackage{amsmath}	
\usepackage{amssymb}	
\let\textbf\relax

\newcommand{\kms}{\ensuremath{\textrm{km}\,\textrm{s}^{-1}}}
\newcommand{\masyr}{\ensuremath{\textrm{mas}\,\textrm{yr}^{-1}}}

\defcitealias{Barba2018}{B19}
\newcommand{\bprp}{\ensuremath{\textrm{G}_\textrm{\textsc{bp}}-\textrm{G}_\textrm{\textsc{rp}}}}

\title[The orbit of FSR1758]{The retrograde orbit of the globular cluster FSR1758 revealed with \textit{Gaia} DR2}

\author[Simpson]{
Jeffrey D. Simpson$^{1}$\thanks{E-mail: jeffrey.simpson@unsw.edu.au}
\\
$^{1}$School of Physics, UNSW, Sydney, NSW 2052, Australia
}

\date{Accepted 2019 June 18. Received 2019 June 13; in original form 2019 January 28}

\pubyear{2019}

\begin{document}
\label{firstpage}
\pagerange{\pageref{firstpage}--\pageref{lastpage}}
\maketitle

\begin{abstract}
We report the first radial velocity measurements of the recently identified globular cluster FSR1758. From the \textbf{two member} stars with radial velocities from the \textit{Gaia} Radial Velocity Spectrograph reported in \textit{Gaia} DR2, we find FSR1758 has a radial velocity of $227\pm1$\kms. \textbf{We also find potential extra-tidal star lost from the cluster in the surrounding 1~deg.} Combined with \textit{Gaia} proper motions and photometric distance estimates, this shows that FSR1758 is on a relatively retrograde, radial orbit with an pericentre of $3.8_{-0.9}^{+0.9}$~kpc, an apocentre of $16_{-5}^{+8}$~kpc, and eccentricity of $0.62_{-0.04}^{+0.05}$. Although it is currently at a Galactocentric distance of $3.8_{-0.9}^{+0.9}$~kpc --- at the edge of the bulge --- it is an intruder from the halo. We investigate whether a reported `halo' of stars around FSR1758 is related to the cluster, and find that most of these stars are likely foreground dwarf stars. We conclude that FSR1758 is not a dwarf galaxy, but rather a globular cluster.
\end{abstract}




\section{Introduction} \label{sec:intro}
The second data release of the \textit{Gaia} mission \citep{GaiaCollaboration:2016cu,GaiaCollaboration:2018io} has revolutionized our view of the Milky Way Galaxy. Among its many results, it has proven very useful for investigating purported stellar clusters, in many cases confirming their existence \citep[e.g.,][]{Simpson:2017ex,Soubiran2018,Cantat-Gaudin2018}, but in some cases showing they are not real physical associations \citep[e.g.,][]{Kos:2018we}. The precise proper motions are especially helpful in the Milky Way bulge, where imaging surveys are hampered by the large and differential reddening and extinction, resulting in clusters hidden from easy view. Its radial velocity measurements, though only available for the brighter cohort of stars, enable us to calculate the 3D motion of many clusters.

Recently \citet[][hereafter \citetalias{Barba2018}]{Barba2018} used data from \textit{Gaia} DR2 and the DECam Plane Survey \citep[DECaPS;][]{Schlafly2018} to present a physical characterization of the large, massive stellar grouping FSR1758. They found the stellar cluster to be curiously large (on the order of the size of $\omega$~Cen), and claimed that it could potentially be the core of a dwarf galaxy based upon a halo of common proper motion stars in the surrounding region. As noted by \citetalias{Barba2018}, the key data missing were spectroscopic observations of the cluster. These data are crucial for deriving an orbit, confirming the photometric metallicity, and understanding the `halo' of common proper motion stars.

In this \textbf{work} we report the radial velocities of \textbf{two} members \textbf{and one extra-tidal member} of FSR1758 measured with the \textit{Gaia} Radial Velocity Spectrograph \citep[RVS;][]{Cropper2018,GaiaCollaboration:2018fx} (Sec.\ \ref{sec:data}), which are used to calculate an orbit for FSR1758 (Sec.\ \ref{sec:orbit}). We also examine the `halo' of common proper motion stars around FSR1758 that led \citetalias{Barba2018} to propose that FSR1758 may be a dwarf galaxy (Sec.\  \ref{sec:halo}). We conclude that these stars are foreground field stars and are not associated with the cluster.

\section{Data}\label{sec:data}
\begin{figure*}
\includegraphics[width=\textwidth]{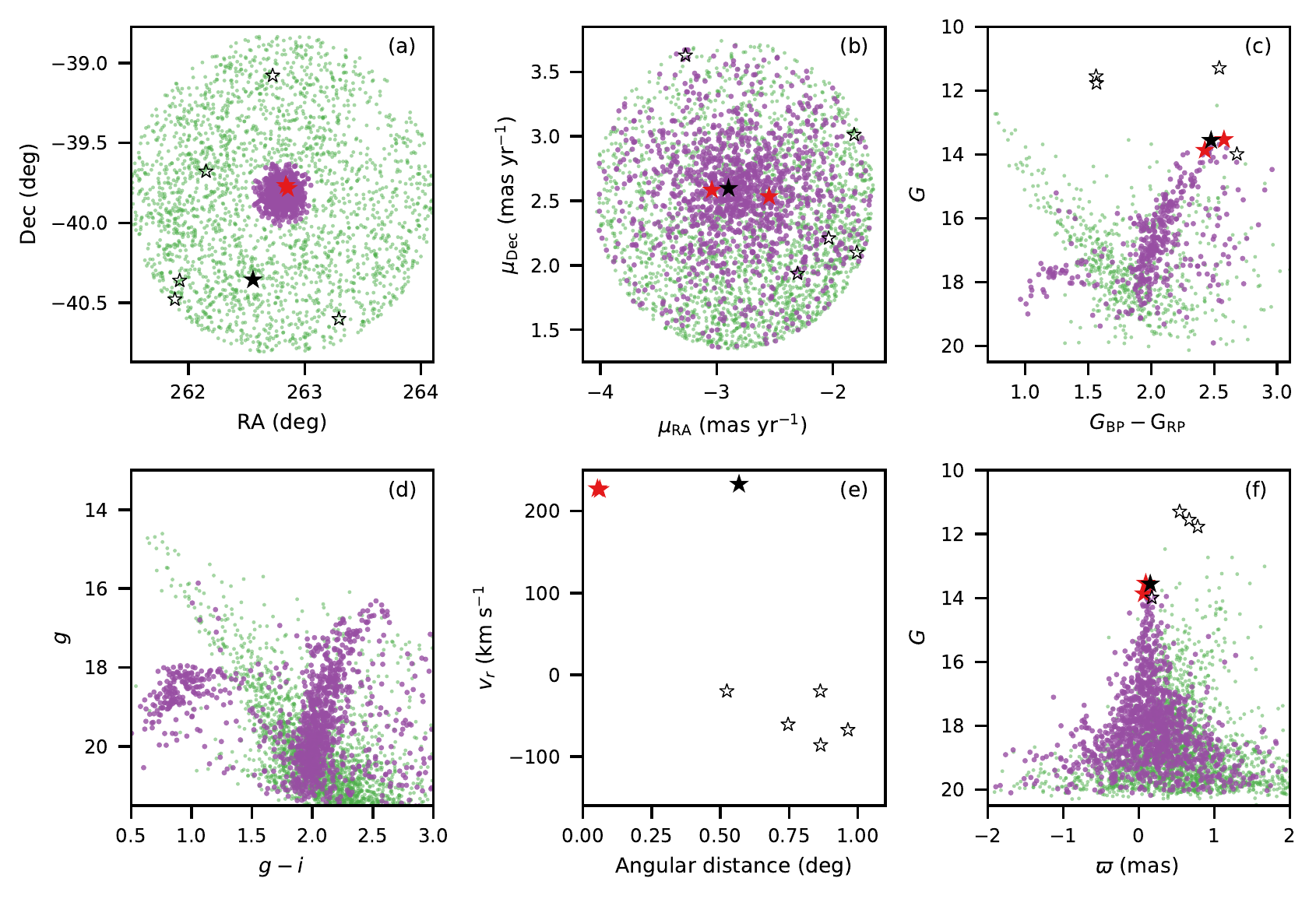}
\caption{Position (a), proper motion (b), color-magnitude diagrams (c,d), radial velocity (e), and parallax (f) distributions of all stars in \textit{Gaia} within 1~deg of FSR1758 and with proper motions within 1.2~\masyr\ of the value found by \citetalias{Barba2018}. The stars have been divided into `cluster' (red stars for those with radial velocity, otherwise purple points) and `field' (black stars for those with radial velocity, otherwise green points) samples based upon their angular distance from the cluster. Filled stars identify the radial velocity members of FSR1758, and the unfilled stars are \textbf{field stars} with radial velocities. There are no stars with radial velocities in (d) due to the bright limit of DECaPS in this region, and \textbf{one of the field stars with a radial velocity is not plotted on panels (c) and (f) because it has $G=8.5$.} The cluster stars show an obvious grouping in proper motion space (panel b), and the colour-magnitude diagram morphology expected for an ancient, metal-poor population (panels c and d). In (f) the distribution of the parallaxes of cluster stars \textbf{is centred about $\varpi\sim0$, with increasing scatter with increasing faintness of the stars. The radial velocity members are all found at the apex of this distribution, while most of the field stars are scattered towards larger parallaxes.}}
\label{fig:cmd}
\end{figure*}

\begin{table*}
\centering
\caption{\textbf{All of the stars with \textit{Gaia} radial velocities that meet the proper motion selection criteria. The first two entries are the FSR1758 members. The parallaxes have had 0.052~mas corrections applied \citep{Zinn2018,Leung2019}.}}
\label{table:star_params}
\begin{tabular}{rrrrrrrrrrrrrrr}
\hline
source\_id & RA & Dec & $\varpi$ & $v_r$ & \bprp & $G$ & $\mu_\mathrm{RA}$ & $\mu_\mathrm{Dec}$ & ang. dist \\
 & (deg) & (deg) & (mas) & (\kms) &  &  & (\masyr) & (\masyr) & (deg) \\
\hline
5961661825166384256 & $262.855$ & $-39.785$ & $0.094$ & $227.49$ & $2.58$ & $13.54$ & $-2.55$ & $2.53$ & $0.05$ \\
5961664784449145344 & $262.838$ & $-39.766$ & $0.068$ & $226.51$ & $2.43$ & $13.87$ & $-3.04$ & $2.58$ & $0.06$ \\
5960133370929943296 & $262.555$ & $-40.357$ & $0.153$ & $233.01$ & $2.47$ & $13.56$ & $-2.90$ & $2.60$ & $0.57$ \\
5960228611762486016 & $262.151$ & $-39.678$ & $0.668$ & $-20.06$ & $1.56$ & $11.55$ & $-2.04$ & $2.21$ & $0.52$ \\
5959364812304283136 & $263.295$ & $-40.602$ & $0.174$ & $-86.11$ & $2.68$ & $13.99$ & $-1.79$ & $2.10$ & $0.86$ \\
5960164260340007424 & $261.883$ & $-40.479$ & $0.541$ & $-67.44$ & $2.54$ & $11.29$ & $-1.82$ & $3.01$ & $0.96$ \\
5960169311221161472 & $261.925$ & $-40.362$ & $0.892$ & $-20.02$ & $2.60$ & $8.48$ & $-3.27$ & $3.63$ & $0.86$ \\
5961768374719872768 & $262.724$ & $-39.078$ & $0.781$ & $-60.81$ & $1.56$ & $11.77$ & $-2.30$ & $1.94$ & $0.75$ \\
\hline
\end{tabular}
\end{table*}

The primary source of data for this work was \textit{Gaia} DR2, combined with DECaPS. We searched for \textit{Gaia} DR2 targets within 1.0~deg of FSR1758 ($[\alpha,\delta]=[262.806^\circ,-39.822^\circ]$) which returned a catalogue of over 1.5 million stars\footnote{The data and analysis code is available at \url{https://doi.org/10.5281/zenodo.2550945}.}. These were positionally cross-matched to the DECaPS catalogue using \textsc{topcat} \citep{Taylor:2005wx,Taylor:2006wv}. The bright limit for the DECaPS photometry is $G\sim14$, which meant that not all of the \textit{Gaia} targets have DECaPS photometry.

\textbf{For regions of the sky with high stellar density (i.e., the Galactic bulge, globular clusters), the results from \textit{Gaia} DR2 can suffer from source confusion and blending that could affect the astrometry \citep[e.g.,][]{Lindegren:2018gy}, photometry \citep[e.g.,][]{Babusiaux:2018di}, and radial velocity measurements \citep[e.g.,][]{Boubert2019}. To minimize the effect of this on the data, we required the following astrometric quality criteria\footnote{These criteria do remove likely members from consideration, and we note in particular \textit{Gaia} \texttt{source\_id} 5961664612650449536, which meets the proper motion, radial velocity, and photometric criteria discussed below for cluster membership, but has $\texttt{ruwe}=1.7$ and $\texttt{rv\_nb\_transits}=2$.}:}
\begin{align}
	\texttt{ruwe} < & 1.4,
\end{align}
\textbf{where \texttt{ruwe} is the re-normalised Unit Weight Error \citep[defined in][]{Lindegren2018}. To assess the radial velocity quality, we required}
\begin{align}
\texttt{rv\_nb\_transits} \geq & 5.
\end{align}
\textbf{For the \textit{Gaia} colour-magnitude diagram (CMD), we required that \begin{equation}\label{eq:good_photom}
	1.0 + 0.015(\bprp)^2<\texttt{CE}<1.3 + 0.06(\bprp)^2,
\end{equation}
where \texttt{CE} is the \texttt{phot\_bp\_rp\_excess\_factor} \citep{Babusiaux:2018di}. This removes stars likely suffering from blends in their blue and red photometric measurements.}

Visually, the cluster is not obvious as an over-density on the sky, but the cluster has a distinct proper motion \citep{Cantat-Gaudin2018, Barba2018}. Stars with proper motions within 1.2~\masyr\ of $(\mu_\mathrm{RA},\mu_\mathrm{Dec})=(-2.85,2.55)$~\masyr\ were extracted from the catalogue. Those within 0.2~deg of the cluster position were selected as `cluster' stars, and those outside that as a `field' sample. This `cluster' sample and the surrounding field are \textbf{presented} in Fig.\ \ref{fig:cmd} \textbf{showing} their position on the sky, proper motions, CMD, radial velocities and parallaxes. The proper motions of the cluster sample are clearly clumped, and the CMD shows the red giant branch (RGB) and horizontal branch (HB) morphology of an ancient, metal-poor stellar population in both the \textit{Gaia} and DECaPS filters.

Of the \textbf{1291} likely cluster stars within 0.2~deg, \textbf{two} stars had radial velocities ($v_r$) measured by the \textit{Gaia} RVS: \textbf{226.5~\kms and 227.5~\kms.} (\textbf{details are in Table \ref{table:star_params} and they are also shown as red} filled stars on Fig.\ \ref{fig:cmd}). \textbf{These two} stars are \textbf{both} found at the tip of the giant branch of FSR1758 (Fig.\ \ref{fig:cmd}c). \textbf{In Fig.\ \ref{fig:cmd}f we show the parallax of the stars against their apparent magnitude \citep[applying a $+0.052$~mas zero-point correction;][]{Zinn2018,Leung2019}. The sample of likely members of FSR1758 have a distribution centred on $\varpi\sim0$~mas, as would be expected for stars at a distance of 11~kpc. The two members with radial velocities are found at the apex of this distribution. Their parallaxes do not provide any information to discount these two stars as members of FSR1758.} We conclude that these \textbf{two} stars are members of FSR1758, and that the radial velocity of the cluster is $v_r=227\pm1$~\kms.

\citetalias{Barba2018} estimated the tidal radius to be $0.78\pm0.22$~deg, so we extended the search for possible members beyond the initial $0.2$~deg sample. Within the surrounding 1~deg, there is one additional star with a large radial velocity ($v_r=233.0~\kms$; black filled star on Fig.\ \ref{fig:cmd})), and it is, like the \textbf{two} stars above, compatible with the proper motion, parallax, and photometry of the cluster. The difference between 233~\kms\ and 227~\kms\ is consistent with the dispersion seen in other globular clusters \cite[e.g., $\omega$~Cen has a velocity dispersion of $\sim10~\kms$;][]{Johnson:2010fs}.

The \textit{Gaia} RVS spectra were not made public in DR2, so we cannot estimate a metallicity for FSR1758. There does not appear to have been any serendipitous spectroscopic observations of the cluster. Unfortunately it is outside the footprints of RAVE \citep{Kunder:2017gp}, GALAH \citep{DeSilva:2015gr,Buder2018}, APOGEE-2 \citep{Zasowski:2017jy} and other bulge spectroscopic surveys \citep[e.g.,][]{Freeman2013,Zoccali2014,Howes:2016ji}.

\section{Orbit}\label{sec:orbit}

\begin{figure*}
	\includegraphics[width=\textwidth]{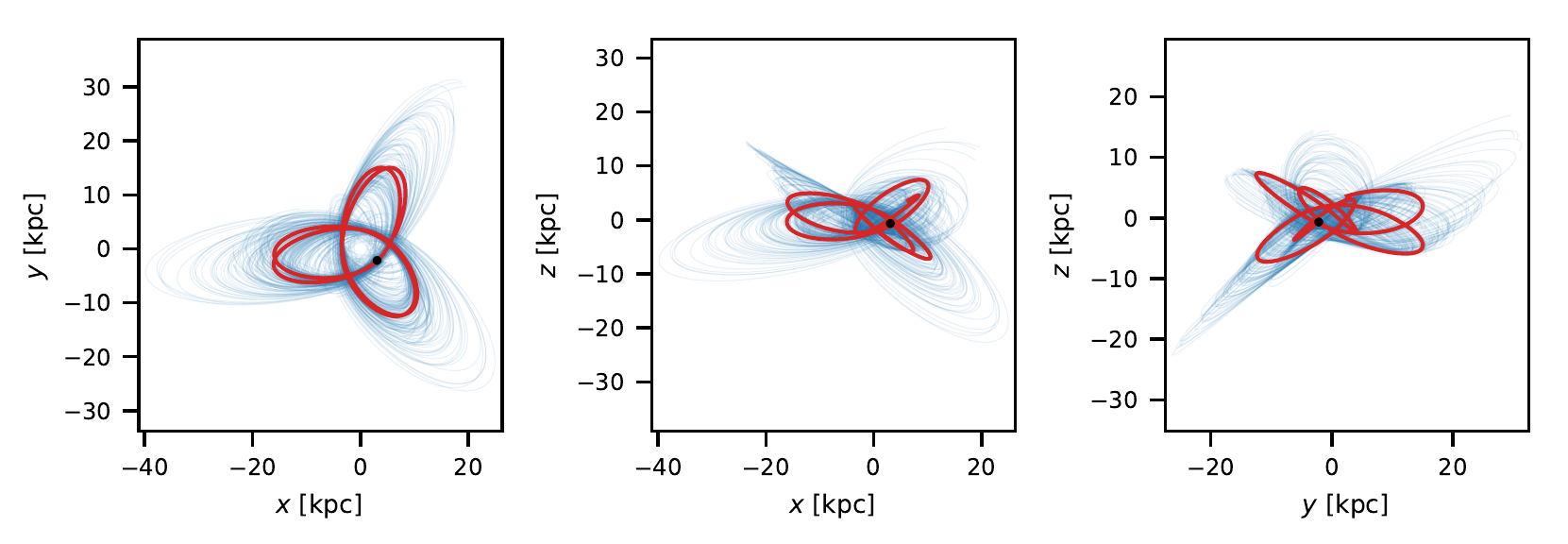}
    \caption{The previous 1.25~Gyr of the orbit of FSR1758 projected into Cartesian space centred on the Galactic Centre. The red line is the orbit from the nominal values of the phase space coordinates, and the blue lines show 100 orbits randomly sampling the error distributions of the input parameters. The black dot shows the observed position of FSR1758.}
    \label{fig:orbit}
\end{figure*}

\begin{figure*}
	\includegraphics[width=\textwidth]{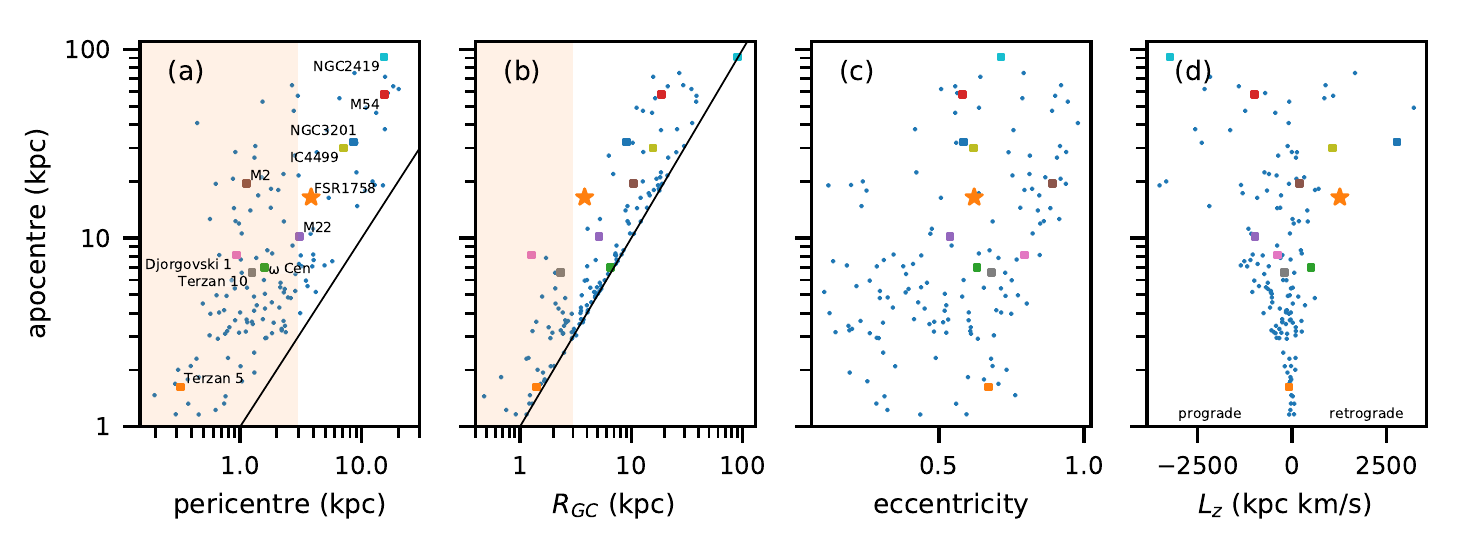}
    \caption{Comparison of the orbit of FSR1758 to other globular clusters. In panels (a) and (b) the shaded region indicates a distance of 3~kpc (i.e., the radius of the Milky Way bulge), and the straight lines show the one-to-one line. For panel (a) a cluster on this line would have a circular orbit, and in panel (b) a cluster on this line is being observed at apocentre. As well as FSR1758, a number of clusters of interest are highlighted. Djorgovski~1 and Terzan~10 were recently found by \citet{Ortolani2019} to be halo interlopers in the bulge. M54, M2, M22, $\omega$~Cen have been found to be multi-metallicity clusters, and could therefore be stripped dwarf galaxy cores. NGC2419, NGC3201, IC4499 have large angular momentum orbits.}
    \label{fig:orbit_comparison.pdf}
\end{figure*}

It is now possible to calculate the orbit of FSR1758. We used \textsc{gala} \citep[version 0.3;][]{Price-Whelan2017a,Price-Whelan2018b}, with the default potential \texttt{MilkyWayPotential}. This is a simple mass-model for the Milky Way consisting of a spherical nucleus and bulge, a Miyamoto-Nagai disk, and a spherical NFW dark matter halo. The parameters of this model are set to match the circular velocity profile and disk properties of \citet{Bovy:2015gg}. We place the Sun at a Galactocentric distance of $R_\mathrm{GC}=8.2$~kpc and a height above the plane of 25~pc \citep{BlandHawthorn:2016iq}. The Sun's velocity is taken to be $(U_\odot,V_\odot,W_\odot)=(11.0,248.0,7.25)~\kms$ \citep{Schonrich2012}. The position and velocity of FSR1758 were $(\alpha,\delta,D_\odot,\mu_\mathrm{RA},\mu_\mathrm{Dec},v_r)=(262.806^\circ,-39.822^\circ,11.5\pm1.0~\mathrm{kpc},-2.85\pm0.1~\mathrm{\masyr},2.55\pm0.1~\mathrm{\masyr},227\pm1~\mathrm{\kms})$. Errors in the calculated orbital parameters were estimated by taking 1000 samples of the error distributions and finding the 16th and 84th percentiles of the given results.

Fig.\ \ref{fig:orbit} shows the previous 1.25~Gyr of its orbit (red line), as well as 100 other possible orbits over the same time interval created by sampling the error distributions (blue lines). Fig.\ \ref{fig:orbit_comparison.pdf} shows how the orbit of FSR17578 compares to the orbit of other known globular clusters, with input data from \citet{Vasiliev:2018uf}. The orbits were calculated in the same manner as FSR1758 (i.e., using \texttt{gala} with the input values for the Milky Way and Sun as described above).

FSR1758 is found to have a radial, retrograde orbit, with a pericentre of $3.8_{-0.9}^{+0.9}$~kpc, an apocentre of $16_{-5}^{+8}$~kpc, and eccentricity of $0.62_{-0.04}^{+0.05}$.  FSR1758 is currently located at a Galactocentric distance of $R_\mathrm{GC} = 3.8_{-0.9}^{+0.9}$~kpc, placing it at the edge of the Milky Way bulge \citep[clusters within $\sim3$~kpc of the Galactic centre tend to be classified as `bulge' clusters;][]{Barbuy2018}. But we find that it is not a cluster that lives in the inner Galaxy like, e.g., Terzan~5. Instead it is a halo intruder into the inner Galaxy, like Djorgovski~1 and Terzan~10 \citep{Ortolani2019}. We have caught FSR1758 near the pericentre of its orbit.

\section{Is FSR1758 a dwarf galaxy or globular cluster?}\label{sec:halo}

\begin{figure*}
	\includegraphics[width=\textwidth]{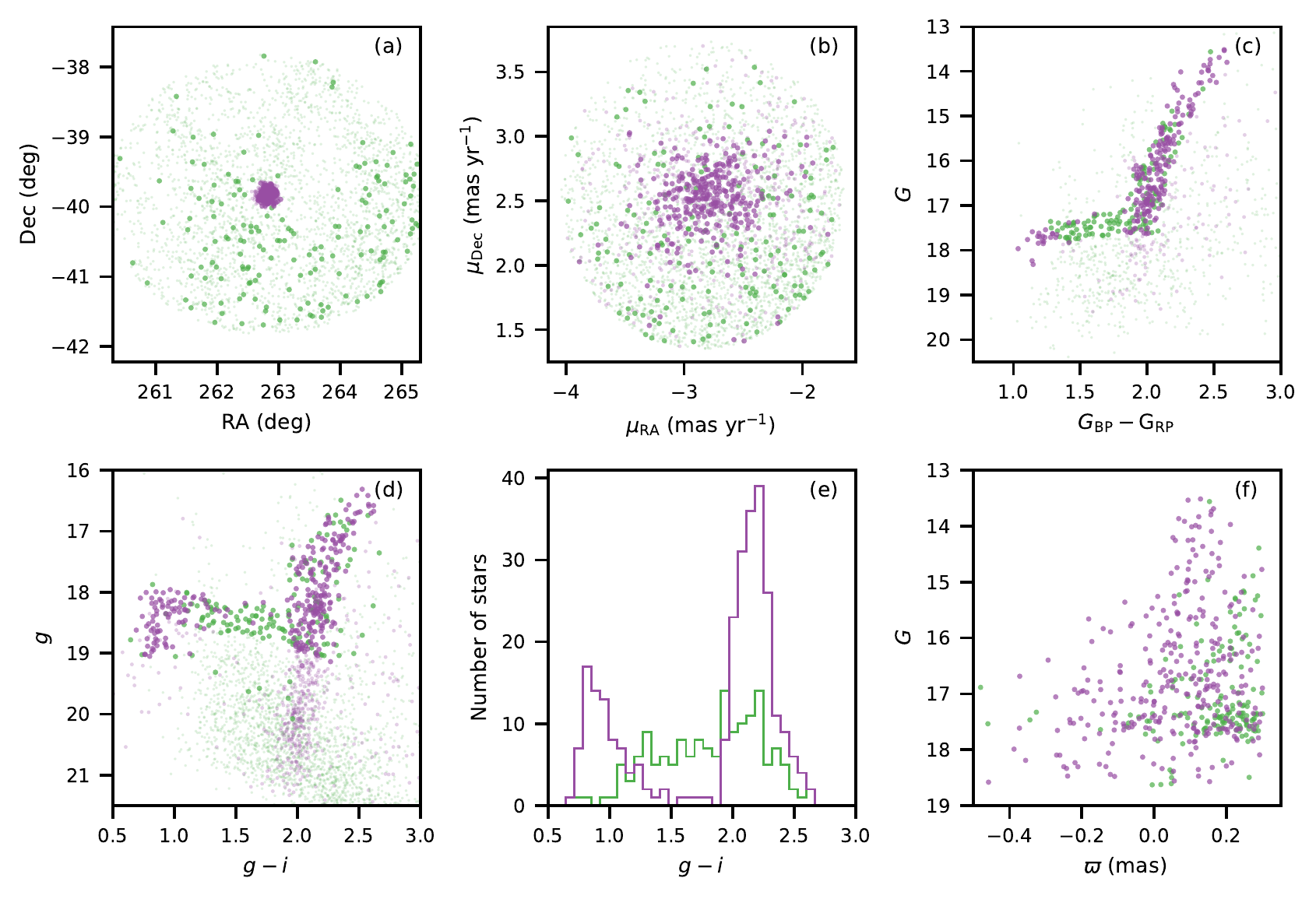}
    \caption{In \citetalias{Barba2018}, they found a halo of common proper motion stars within 2~deg of the cluster. We repeat their analysis, selecting stars along a locus formed by the RGB and HB in the \textit{Gaia} filters within 2~deg of the cluster, with proper motions like the cluster, and with parallaxes $<0.3$~mas. As in Fig.\ \ref{fig:cmd} they are divided into the `cluster' and `halo' sample based upon their angular distance from the cluster. Considering the distributions of proper motion, colour and parallax, we conclude that the majority of this `halo' are field stars that are coincidently share the proper motion of FSR1758.}
    \label{fig:not_a_halo}
\end{figure*}

One of the unresolved questions of \citetalias{Barba2018} was whether FSR1758 is a dwarf galaxy or globular cluster. The orbital properties of FSR1758 do not distinguish it from other globular clusters, or proposed stripped dwarf galaxy cores (e.g., M54, $\omega$~Cen). We find that FSR1758 has quite a retrograde orbit (i.e., large $L_z$) compared to most other clusters. Clusters with retrograde orbits are usually classified as being accreted by the Milky Way rather than forming in situ \citep[e.g.,][]{Kruijssen2018}. The large retrograde component of the inner halo of the Milky Way found in \textit{Gaia} DR2 has been associated with a number of globular clusters, including $\omega$~Cen \citep{Helmi:2018wy}, but these have smaller $L_z$ than FSR1758, so we do not associated FSR1758 with \textit{Gaia}-Enceladus. An intriguing future direction of research, beyond the scope of this work, is to look if any of the large number of stellar streams that are being found in \textit{Gaia} and other surveys are related to FSR1758 \citep[e.g.,][]{Shipp2018,Ibata:2018wn,Malhan:2018ko,Ibata2019}.

Part of the reasoning for the claim in \citetalias{Barba2018} that that FSR1758 could be a dwarf galaxy is a large `halo' of common proper motion stars around the cluster, which they interpreted as possibly being tidal debris or that FSR1758 was the nucleus of a dwarf galaxy. In Fig.\ \ref{fig:not_a_halo}, we repeat their analysis: stars along a locus formed by the RGB and HB within 2~deg of the cluster, with parallax $<0.3$~mas, and proper motions within $1.2\masyr$ of the nominal cluster value. We divide this selection of stars into a `cluster' and 'halo' sample based upon angular distance from the cluster (cut at 0.2~deg).

We conclude that the majority of this halo of stars are not truly associated with FSR1758, but are instead mostly foreground field dwarf stars. There is a clear colour gap between the HB and RGB, especially obvious in the DECaPS photometry (Fig.\ \ref{fig:not_a_halo}d). Only 11 per cent (\textbf{37/328}) of the `cluster' stars are found in the colour range $1.1<g-i<2.0$,  while for the `halo' sample it is \textbf{57} per cent (\textbf{110/194}). This gap corresponds to the field dwarf sequence along the line of sight (see also Fig.\ \ref{fig:cmd}c,d). FSR1758 shows an obvious blue horizontal branch, but almost none of the halo sample is found in this region of the CMD. The halo sample is not clumped in proper motion space like the cluster stars (Fig.\ \ref{fig:not_a_halo}b). \textbf{Considering the parallax distributions, as also shown in Fig.\ \ref{fig:cmd}f, the cluster sample is distributed about a common value with increasing dispersion with increasing faintness. The `halo' sample on the other hand tends to be preferentially found at larger parallaxes ($\varpi\sim0.2$);} it is simply the low parallax selection of the broader field parallax distribution.

It is possible that a few of these common proper motion stars are extra-tidal stars, as these are known around other globular clusters \citep[e.g.,][]{Simpson:2017be}, but the majority are unlikely to be related. We do not discount, with its retrograde orbit, the idea that FSR1758 could be the stripped core of a dwarf galaxy (e.g., like $\omega$~Cen, M54) and it should be a high priority to observe this cluster to see if it has a metallicity spread.

\section{Conclusion}
We have identified \textbf{two} members of FSR1758 \textbf{and one extra-tidal star} that have radial velocities measured by the \textit{Gaia} RVS. Combined with the previously derived information about the cluster from \citetalias{Barba2018}, this allows us to calculate an orbit for the cluster which shows that it is a halo intruder into the inner Galaxy. We conclude that a possible large halo of common proper motion stars around the cluster are in fact likely to be less distant field stars.

We reiterate the words of \citetalias{Barba2018} that ``acid test for this cluster will be to obtain spectra for a number of members''. High resolution spectroscopic observations will be extremely useful, allowing us to know if the cluster has a metallicity spread. Given its retrograde orbit, a metallicity spread would indicate that FSR1758 is in fact a stripped dwarf galaxy core. With only \textbf{two} stars, it is not possible to make any meaningful comments about the radial velocity dispersion, and therefore the mass-to-light ratio of FSR1758. Where does it sit on the mass-metallicity relationship \citep{Kirby2013}? If it is simply a globular cluster (i.e., no metallicity spread), then it is still an important object, representing the remnants of a dwarf galaxy system that has now been accreted by the Milky Way.

\section*{Acknowledgements}

JDS acknowledges the support of the Australian Research Council through Discovery Project grant DP180101791. \textbf{He thanks the anonymous referee for their helpful comments, especially with regards to Gaia reliability and quality in crowded regions of the sky.}

The following software and programming languages made this research possible: \textsc{astropy} \citep[v3.1;][]{TheAstropyCollaboration:2018ti,TheAstropyCollaboration2018}, a community-developed core Python package for Astronomy; \textsc{gala} \citep[v3.0;][]{Price-Whelan2017a,Price-Whelan2018b}; \textsc{pandas} \citep[v0.20.3;][]{McKinney:2010un}; \textsc{seaborn} \citep[v0.8.1;][]{Waskom2018}; Tool for OPerations on Catalogues And Tables \citep[\textsc{topcat}, v4.5;][]{Taylor:2005wx,Taylor:2006wv}

This work has made use of data from the European Space Agency (ESA) mission {\it Gaia} (\url{https://www.cosmos.esa.int/gaia}), processed by the {\it Gaia} Data Processing and Analysis Consortium (DPAC, \url{https://www.cosmos.esa.int/web/gaia/dpac/consortium}). Funding for the DPAC has been provided by national institutions, in particular the institutions participating in the {\it Gaia} Multilateral Agreement.






\bsp	
\label{lastpage}
\end{document}